\documentclass[letter]{aa} 

\usepackage{natbib}
\bibpunct{(}{)}{;}{a}{}{,} 
\usepackage{graphicx}
\usepackage{txfonts}
\begin{document}

   \title{Gaia 8: Discovery of a star cluster containing beta Lyrae
   }

   \subtitle{and of a larger old (extinct) star formation complex surrounding it}

   \author{U. Bastian
          \inst{1}
          }

   \institute{Zentrum f\"ur Astronomie (Center for Astronomy), Heidelberg University,
              M\"onchhofstr. 14, D-69120 Heidelberg\\
              \email{bastian@ari.uni-heidelberg.de}
             }

   \date{Received August 28, 2019; accepted xxx yy, 2019}

 
  \abstract
   {}
   {To try and clarify whether there are stars in the neighbourhood of $\beta$ Lyrae which are physically connected to the famous prototype eclipsing binary.}
   {Scrutinize the astrometric and photometric data (from Hipparcos and Gaia DR2) of stars in a limited sky field around $\beta$~Lyrae.}
   {A star cluster of about 100 members is discovered, nicely centered on $\beta$~Lyrae in 5 dimensions. There are clear indications that the newly discovered cluster belongs to a larger old (extinct) star formation complex, including two already known star clusters and an extended strewn field of stars at practically the same distance and space velocity.}
   {This discovery opens up the possibility of determining a precise distance and a precise age for $\beta$ Lyrae, using the combined properties of the cluster members. This in turn may provide an important asset for an improved astrophysical interpretation of the model binary $\beta$ Lyrae and its kin. --- The larger old star formation complex is of scientific interest in its own right.}


   \maketitle
%

\section{Introduction}

The bright eclising binary $\beta$ Lyrae is an object of considerable historical and astrophysical interest. The SIMBAD database bears witness of its scientific importance by listing almost a thousand\footnote{Precisely 878 as of August 11, 2019.} literature references. The star is the prototype of evolved semi-detached binaries with strong mutual gravitational deformation of the partners, with mass transfer, significant period evolution and occasionally an accretion disk around the receiving partner.   
 
There is a long-running, low-level discussion whether some members of a half-dozen group of stars of apparent magnitudes 7--11 around $\beta$ Lyrae might be physically connected to the massive binary, see e.g.\,\citet{Abt1962} and \citet{Abt1976}. Using the astrometry and photometry of Gaia DR2, the second data release of ESA's astrometry mission \citep{Gaia2018,Lindegren2018}, this question could be expected to find a final answer. Checking the relevant data in DR2 surprisingly revealed an open star cluster of around 100 members to which $\beta$ Lyrae obviously belongs. 
 
It is suggested to denote this cluster as Gaia\,8, in reminiscence of the discovery of the clusters Gaia\,1 (the surprising massive cluster next to Sirius on the sky) and Gaia\,2 (in Perseus at more than 5\,kpc distance) by \citet{Koposov2017}, and in honor of the Gaia astrometry satellite project which made these discoveries possible. 
It is noted that number~8 is given to the new cluster because meanwhile \citet{Torrealba} discovered and named clusters Gaia\,3,4,5,6,7 (still using Gaia DR1).

Sections~2 and~3 of the present paper describe the cluster and its discovery, Section~4 presents its probable connection with an extended, old (extinct) and hitherto unknown star formation complex, and Section~5 briefly discusses the possible scientific usage of $\beta$ Lyrae's membership to the cluster, as well as the possible scientific interest of the surrounding larger stellar aggregate.


\section{Discovery of the cluster}

\begin{figure}
   \centering
      \includegraphics[width=7.6cm]{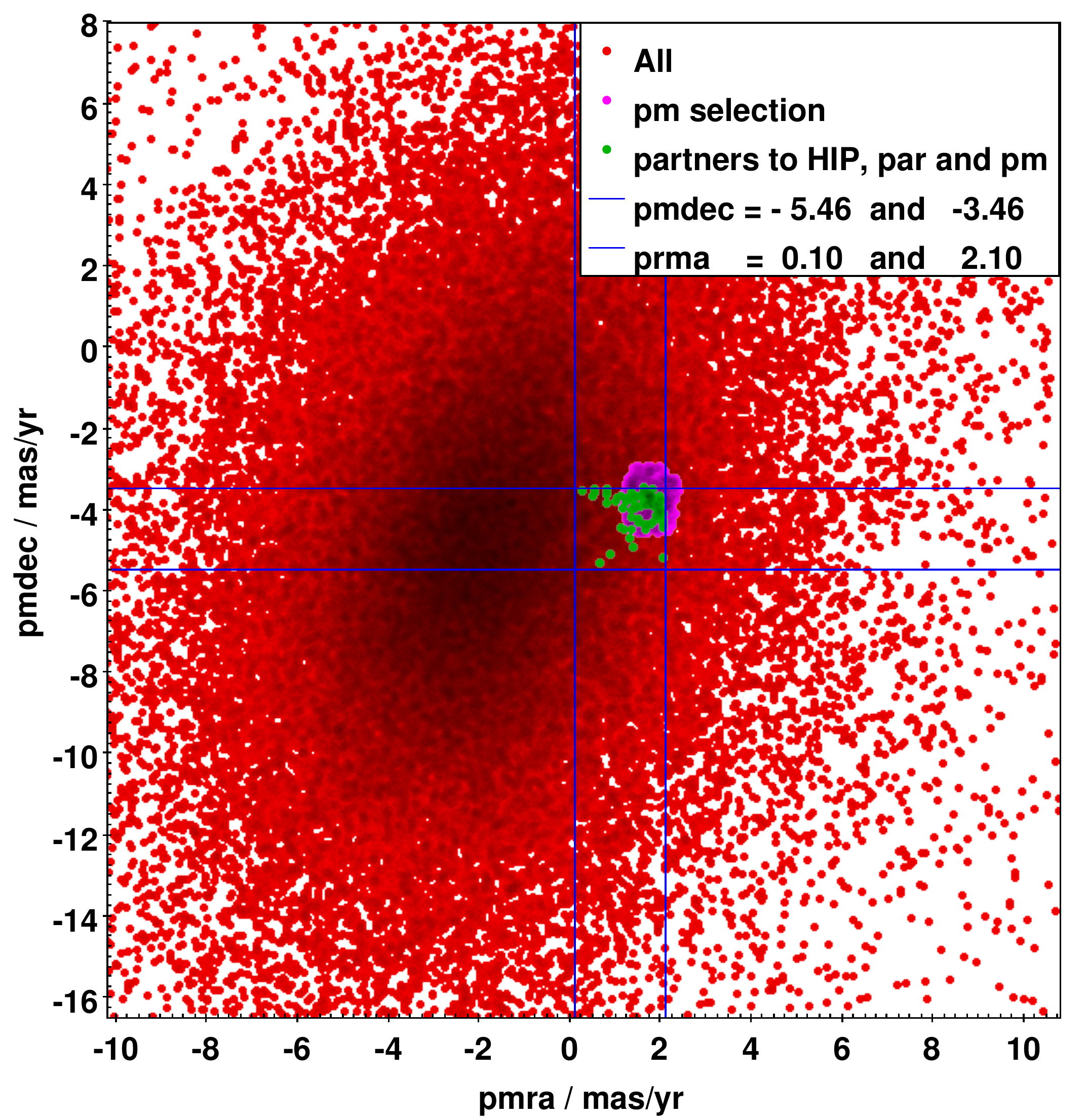}
   \caption{Proper-motion vector diagram of the Gaia DR2 stars in a circle of 1 degree radius around $\beta$ Lyrae. The blue lines and the two sub-samples which led to the discovery of the star cluster are explained in the text. Axes are labelled by the names of the relevant quantities in the public Gaia DR2 tables; units are indicated.}
              \label{pmselection}
    \end{figure}

\begin{figure}
   \centering
   \includegraphics[width=7.6cm]{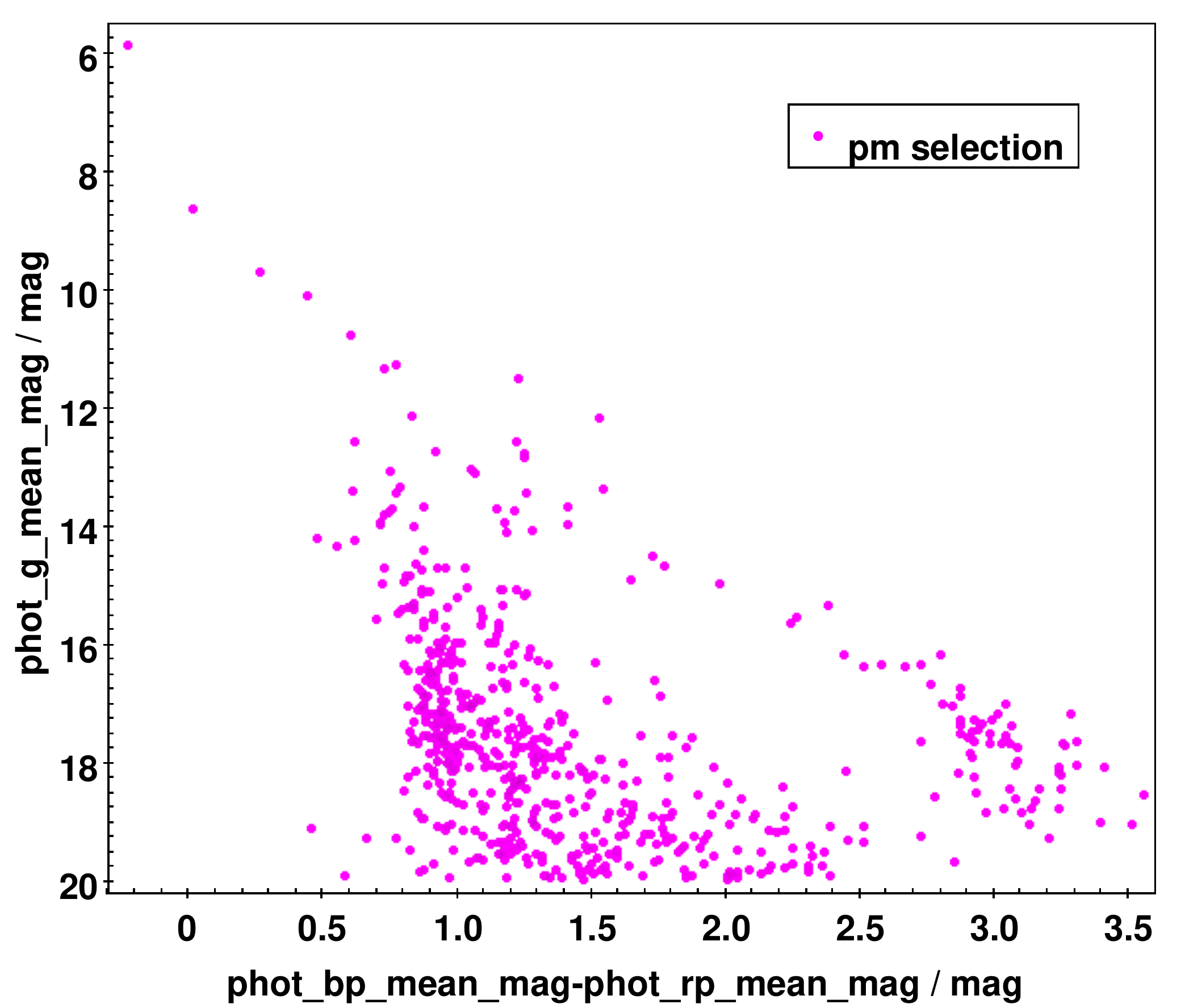}
   \caption{Straight colour-magnitude diagram of the stars in the pink area of the proper-motion vector diagram, Fig.~\ref{pmselection}. The star cluster is indicated by the very clear-cut main sequence of close-by stars, well separated from a set of unrelated galactic background stars accidentally having the same proper motion. Axes are labelled by the names of the relevant quantities in the public Gaia DR2 tables; units are indicated.}
              \label{cmd}%
    \end{figure}

In order to check for objects possibly connected to $\beta$ Lyrae, initially a sky field of 1\,degree radius was retrieved from the public DR2 installation at ARI, Heidelberg (see http://gaia.ari.uni-heidelberg.de/tap.html), the star list being restricted to significant DR2 parallaxes ($\varpi/\sigma_\varpi>8$) and to brightness G<20. The proper-motion vector diagram of this sample is shown as red dots in Fig.~\ref{pmselection}. Further down-selecting on proper motions within $\pm$1\,mas/a around the Hipparcos \citep{Leeuwen2007} motion of $\beta$ Lyrae (blue box in Fig.~\ref{pmselection}) and parallaxes within $\pm$1\,mas around the Hipparcos value for $\beta$ Lyrae led to the sample shown as green dots in Fig.~\ref{pmselection}. Surprisingly, that sample turned out to be clearly concentrated on the sky to the position of $\beta$ Lyrae (not shown, but see Fig.~\ref{skymap}), and even more so on the upper-right corner of the original proper-motion box in Fig.~\ref{pmselection}. The immediately obvious suspicion of a star cluster was checked by dropping the parallax criterion and selecting a re-centered proper-motion sample (pink in Fig.~\ref{pmselection}). This resulted in a highly telling colour-magnitude diagram (Fig.~\ref{cmd}), showing a very well defined main sequence  of nearby stars of equal distances (around 290\,pc), plus a well separated, unrelated galactic background. Due to the small sky field, there are no foreground stars in the selected proper-motion range.           

Hipparcos, rather than DR2 parameters for $\beta$ Lyrae were used here for the following reasons: Gaia can measure this extremely bright star ($G$ magnitude about 3.2--4.4) only in the form of strongly saturated images 
\citep[see] [and references therein] {Gaia2016}. For Gaia DR2 
\citep{Gaia2018,Lindegren2018} these images could not yet be sufficiently calibrated\footnote{Note that Gaia was designed and announced for a bright magnitude limit of $G=5.7$, i.e.~$\beta$ Lyrae is 3--10 times brighter than this limit.} astrometrically. This can be seen from the overall error distribution of stars with G$<$5 in DR2 as well as from some strange individual results for such very bright stars. In addition, quite some very bright stars are missing in the catalogue.

The particular star $\beta$ Lyrae itself is present in DR2, but its G magnitude in DR2 is given as 7.5, its parallax of 1.1\,mas is clearly in discordance with the Hipparcos parallax of 3.4\,mas, and its astrometric quality indicators can best be summarised as ``terrible"\footnote{Specifically (with typical values for stars of magnitude G$<$9 given in parentheses):~astrometric\_gof\_al=237 (ca.~5); astrometric\_excess\_noise=2.1\,mas (mostly zero); parallax\_error=0.34\,mas (0.04\,mas); phot\_bp\_rb\_excess\_factor=4.8 (1.2)}.

\section{Confirmation of the cluster, and of the mem\-ber\-ship of $\beta$ Lyrae}

\begin{figure}
   \centering
   \includegraphics[width=9.0cm]{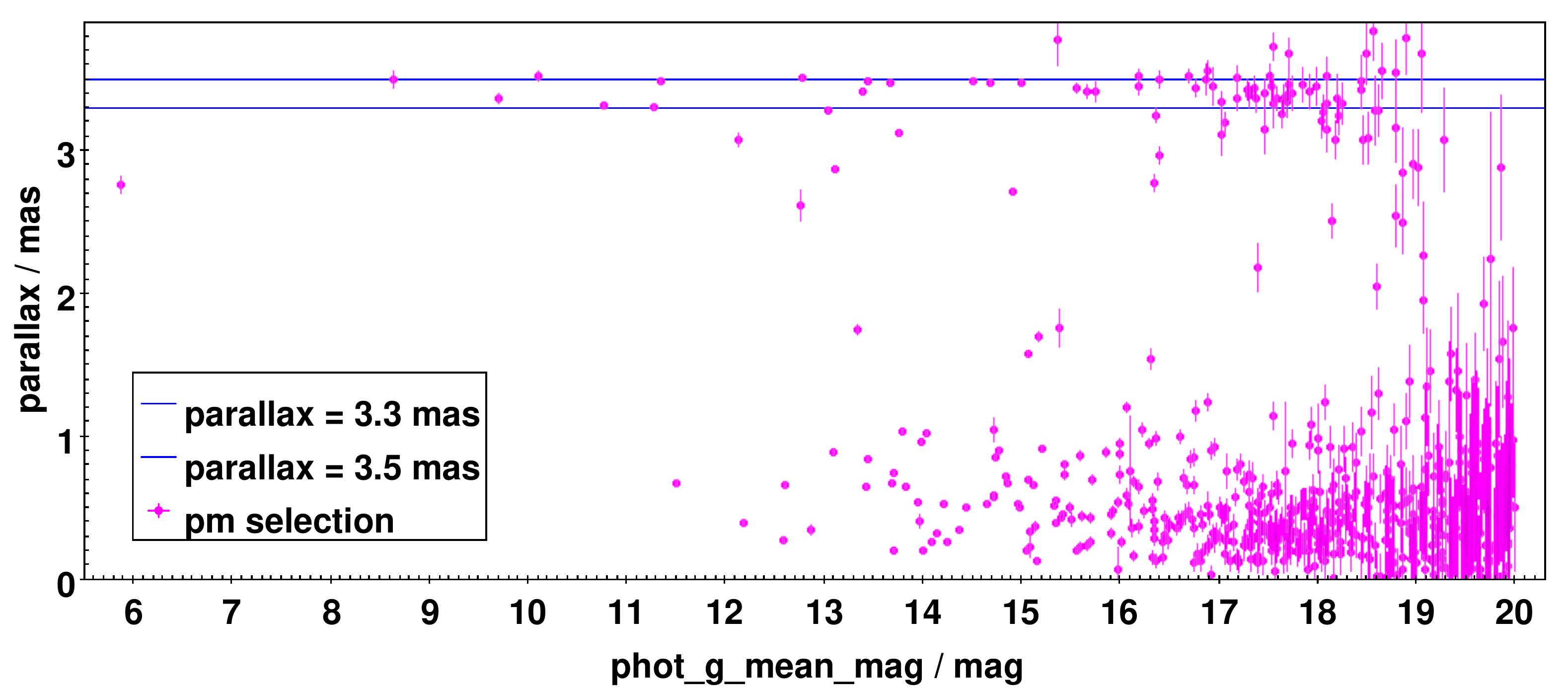}
   \caption{Gaia DR2 parallaxes of the star sample in Fig.~\ref{cmd}, independently confirming the cluster nature of the photometric discovery made via Fig.~\ref{cmd}. The horizontal axis is the Gaia DR2 G~magnitude; axes are labelled by the names of the relevant quantities in the public Gaia DR2 tables; units are indicated. The pink bars indicate the standard errors of the parallaxes, as given in DR2.}
              \label{parallaxes}%
    \end{figure}

A confirmation of the cluster which is completely independent of the proper motions and photometric data used for its discovery can be provided by the distribution of the Gaia DR2 parallaxes. To this purpose, Fig.~\ref{parallaxes} uses the same star sample as ~Fig.~\ref{cmd}. The cluster is beautifully confirmed; the mean parallax of the obvious candidate members is about 3.4\,mas. The scatter of the parallaxes at the bright end of Fig.~\ref{parallaxes} is clearly larger than the indicated DR2 parallax uncertainties. This is not due to measurement errors, but nicely agrees with the putative spatial depth of the cluster: The angular radius of slightly under 1~degree (see Fig.~\ref{skymap} corresponds to 5~pc at a parallax of $\varpi$=3.4\,mas, which translates into $\Delta\varpi$=0.06\,mas in radial direction. Using the individual DR2 parallaxes to derive absolute magnitudes and plotting an ``HRD'', an extremely sharp main sequence is created (see Fig.~\ref{hrd-4clusters}).

The cluster membership of $\beta$ Lyrae is made virtually certain by the following facts:
\vspace {-0.28cm}
\begin{itemize}
\item The cluster is perfectly centered on $\beta$ Lyrae on the sky. This is to be expected if $\beta$ Lyrae is the by-far heaviest member.
\item The proper motion of the cluster is fully consistent with the Hipparcos motion of $\beta$ Lyrae.
\item The mean parallax of the candidate members is practically equal to the Hipparcos value for $\beta$ Lyrae.
\item There are four candidate members with radial velocities given in Gaia DR2. Three of them cluster around -17\,km/s, the fourth one is at +5\,km/s. The General Catalogue of Stellar Radial Velocities (1953) lists $\beta$ Lyrae at -19\,km/s, while \citet{Abt1962} independently found -18.5($\pm$1)\,km/s. 
\end{itemize}

\section{Investigation of the neighbourhood of the cluster}

\begin{figure}
   \centering
   \includegraphics[width=9.0cm]{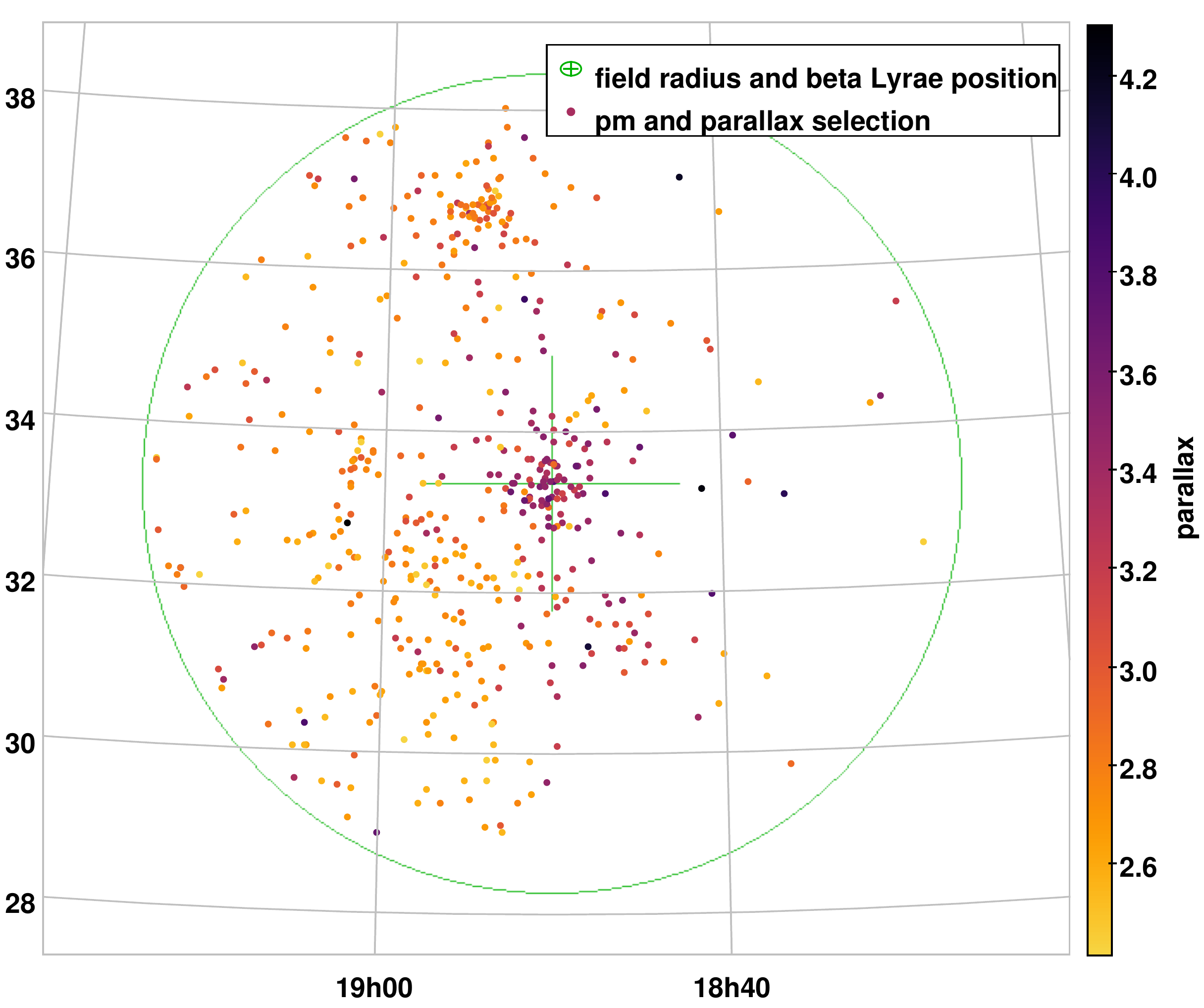}
   \caption{Sky distribution of the stars within 1\,mas/a in proper motion and 1\,mas in parallax around the motion and parallax of the cluster containing $\beta$~Lyrae, and within 5\,degrees angular separation around the position of $\beta$~Lyrae itself (green cross). The individual star symbols are coloured by the DR2 parallaxes. Four stellar aggregates can be seen, as described in the text. }
              \label{skymap}%
    \end{figure}

Closer inspection of Fig.~\ref{parallaxes} reveals that --- compared to a pure field-star distribution --- there is a clear excess of stars at slightly smaller parallax than the $\beta$ Lyrae cluster. Is there another cluster? To investigate this question, an analogous star sample as for ~Fig.~\ref{cmd} was created, but on a larger sky field with 5 degrees of radius, and adding the $\pm$1\,mas parallax criterion around the 3.4\,mas value of the $\beta$ Lyrae cluster. A simple sky map of this sample is given in Fig.~\ref{skymap}. Four features jump to the eye:
\vspace {-0.28cm}
\begin{itemize}
\item The prominent cluster Gaia\,8, perfectly centered on the position of $\beta$ Lyrae, with parallaxes around 3.4\,mas (violet-brownish colours of the plotted dots).
\item Another prominent cluster near the top of the map, with parallaxes around 2.7\,mas (orange-yellowish colours). This is Stephenson\,1 alias $\delta$\,Lyrae\,cluster.
\item A less prominent cluster about 2\,degrees east of $\beta$ Lyrae, also with parallaxes around 2.7\,mas (orange-yellowish colours). This is KPR2005\,100 alias ASCC\,100, see \citet{kharch2005}, \citet{Kharchenko2007}.
\item An extended strewn field of stars southeast of $\beta$ Lyrae and possibly overlapping with Gaia\,8, again with parallaxes around 2.7\,mas and slightly smaller. This is a real stellar agglomeration: From an equivalent skymap avoiding the parallax selection (not shown) it can be definitely excluded that the enhanced density of selected stars in the bottom left part of Fig.~\ref{skymap} is due to interstellar extinction at the right-hand half of the map, or due to the lower galactic latitude at the bottom left of the map. 
\end{itemize}

\begin{figure}
   \centering
   \includegraphics[width=9.0cm]{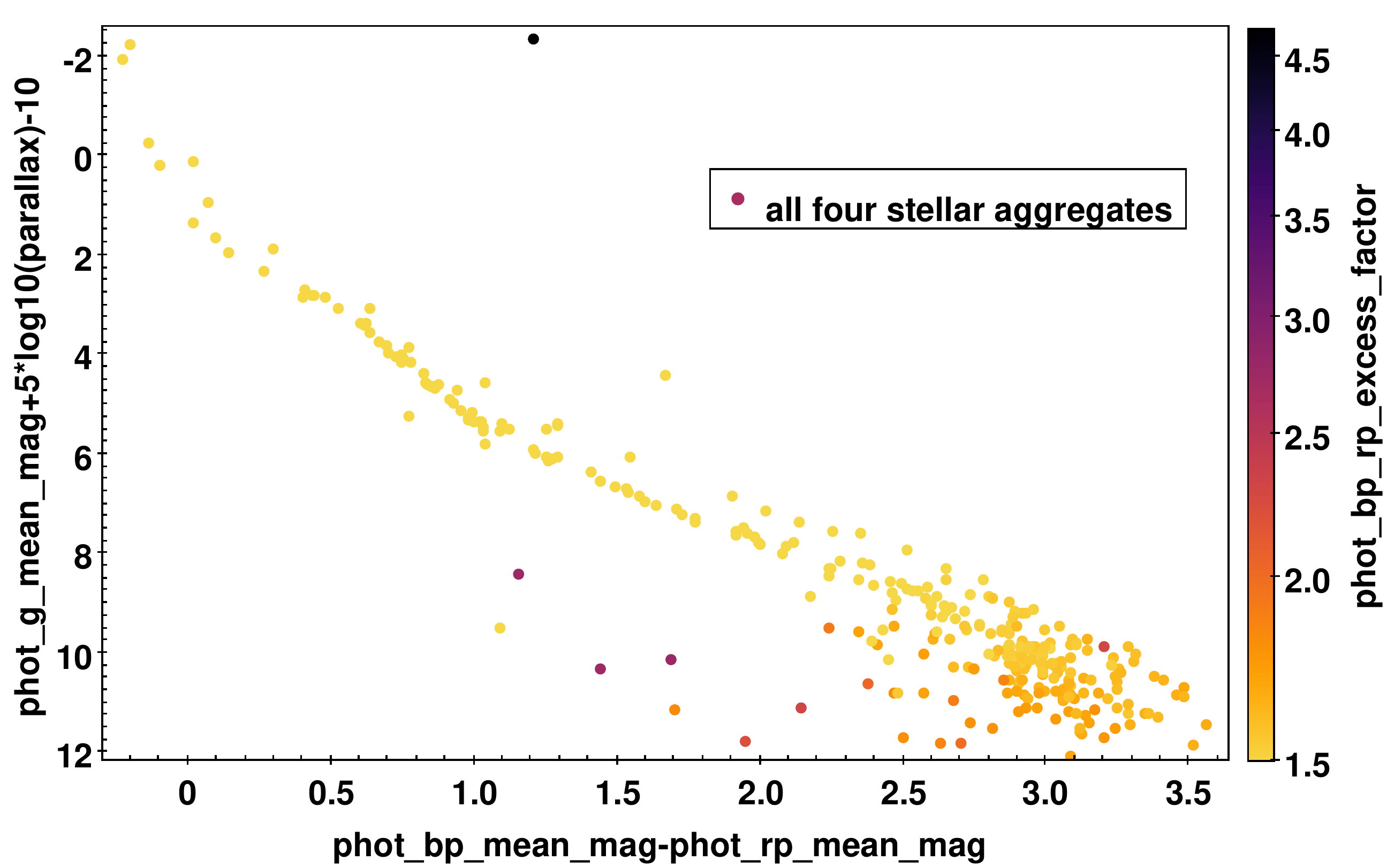}
   \caption{Combined ``HRD'' for the manually selected candidate members of the four stellar groups identified via Fig.~\ref{skymap}. The symbols are coloured according to the so-called BP-RP flux excess from Gaia DR2. For a precise definition and for extensive explanations see the DR2 documentation. In short: High values of this quantity indicate unresolved double stars with disturbed Gaia DR2 colours. This quantity is shown here in order to indicate that the single dark dot at the top is not due to a real giant belonging to the cluster(s), but to a very probably incorrect colour. Similarly, most of the candidate members far left of the main sequence at the bottom are very probably true members, but with disturbed colours.}
              \label{hrd-4clusters}%
    \end{figure}

The immediate impression that these four features might be genealogically connected, i.e.~that they might constitute parts of a larger old (extinct) star formation complex, was checked by analysing manual ad-hoc selections of the candidate members of each feature around their respective centers on Fig.~\ref{skymap}. An intimate connection among them is made virtually certain by the following facts:
\vspace {-0.28cm}
\begin{itemize}
\item All four are located within a very small region of the galactic disk: The largest tangential distance is between the centers of Stephenson 1 and of the strewn field. This is about 30 pc at an assumed parallax of 2.75\,mas. The maximum radial distance is given by the difference in mean parallax between Gaia\,8 and the other three, i.e.~about 70\,pc. Those other three differ by less than 20\,pc in distance, i.e.~less than the on-sky diameter of the strewn field.   
\item All four share practically the same tangential velocity relative to the sun. Although their respective mean tangential velocities differ statistically significantly (due to the high precision of Gaia DR2!), they are all less than 1\,km/s from the one of Gaia\,8. As an aside it is noted that the proper motion of Gaia\,8 nevertheless differs more strongly from the other three, due to the smaller distance.
\item All four share the same radial velocity. Among the 521 stars in Fig.~\ref{skymap} there are 43 with a radial velocity given in Gaia DR2. Among these, 41 are consistent with a uniform velocity of -18\,km/s. They are distributed over all four candidate member selections.
\item A combined ``HRD'' (Fig.~\ref{hrd-4clusters}) for all four selections (using the individual DR2 parallaxes per star to derive absolute G magnitudes) shows an extremely uniform and sharp main sequence, along with a well-defined binary sequence running in parallel. The narrowness of this main sequence points to a similar age and metallicity of the entire complex. It is a pity that no genuine subgiants and giants belong to the complex. Else a precise age could directly be derived from Fig.~\ref{hrd-4clusters}. An age between 30 and 100 Myears is indicated, but this remains to be detailed by a quantitative isochrone fitting (in a follow-up study).
\end{itemize}   

\begin{figure}
   \centering
   \includegraphics[width=9.0cm]{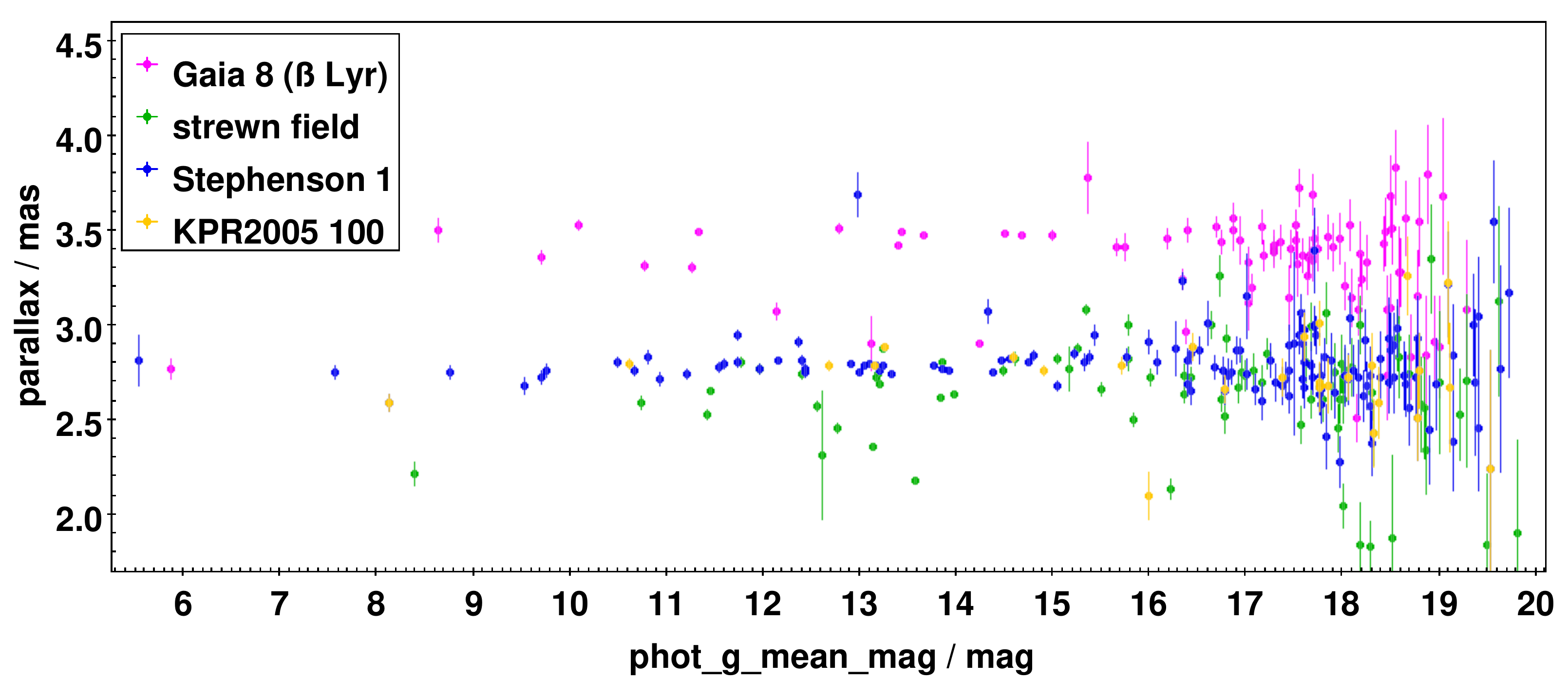}
   \caption{Gaia DR2 parallaxes of candidate members of the three clusters and the strewn field of Fig.~\ref{skymap}. The horizontal axis is the Gaia DR2 G~magnitude; axes are labelled by the names of the relevant quantities in the public Gaia DR2 tables; units are indicated. The bars indicate the standard errors of the parallaxes, as given in DR2. Colours denote tentative membership assignments, see text.}
              \label{parallax-vsGmag-plus-errors-5deg-clusters}%
    \end{figure}

Another informative view of the contents of Fig.~\ref{skymap} is given by Fig.~\ref{parallax-vsGmag-plus-errors-5deg-clusters}. It is an equivalent of Fig.~\ref{parallaxes}, but now for the same star selection as in Fig.~\ref{hrd-4clusters}, and with the individual symbols coloured according to the manual membership selection of the four stellar groups discussed above. It can be seen that the originally ``disturbing'' parallaxes from Fig.~\ref{parallaxes} are actually members of the strewn field superposed on the cluster Gaia\,8 (pink dots in the lower band of symbols), while a few stars assigned to other aggregates (from the purely position-based sample selections) are in fact outlying members of Gaia\,8 (blue and green dots in the upper band of symbols). The latter are also directly apparent in Fig.~\ref{skymap} as loosely scattered dark dots. In addition, Fig.~\ref{parallax-vsGmag-plus-errors-5deg-clusters} nicely shows the difference in distance between Gaia\,8 and the other three stellar groups, as well as the similarity of the distance among these three --- plus the presence of a few stars in the strewn field with larger distances (green dots at the bottom). Some of them may be unrelated background stars, and especially so the single yellow dot at parallax $\simeq$2.1\,mas and G$\simeq$16. 

\section{Discussion}
\label{discussion}

Clearly, the newly found cluster Gaia\,8 as well as the larger stellar aggregate deserve closer investigation. The astrometric part of Gaia DR3 --- giving more precise parallaxes by about a factor~1.25 and more precise motions by almost a factor~2 --- is supposed to be published in less than a year. Therefore it seems reasonable to use that release for this purpose. The main aim of the present short paper is to prompt the collection of supportive detailed observational data on some members in the meantime. Lists of candidate members will be provided by the author on request.

The scientific motivation for deeper investigations will --- on the one hand --- be given by the astrophysical interest in $\beta$\,Lyrae itself. Using precise multi-colour photometry in combination with high-quality spectroscopy of a small number of bright single-star members, or for example asteroseismology of just one or two bright members, can give solid estimates of the age and metallicity of the cluster. This in turn will provide a very useful asset for an improved astrophysical interpretation of the model binary $\beta$~Lyrae and its kin.

On the other hand, the larger stellar aggregate can serve as an interesting study case for the long-term future of extended, diffuse star formation regions, including double clusters etc. 
In fact several such similar groups and complexes of stars and clusters have been identified since 2018 (all from Gaia data), by e.g.~\citet{Cantat-Gaudin}, or \citet{Kounkel}, and possibly by \citet{Franciosini}. There is no doubt that significant progress will be made on this interesting topic in the near future.

Gaia 8, ASCC 100, Stephenson 1, and the sparser strewn field (i.e.~all the stars displayed in Figures~\ref{skymap}, \ref{hrd-4clusters} and~\ref{parallax-vsGmag-plus-errors-5deg-clusters}) apparently form a subgroup of the coeval complex labelled ``Stephenson 1" in Figure~10 of \citet{Kounkel}. 
By assuming that the entire complex shares a common age, they estimate log(age)$\simeq$7.4 (25~Myear). This can surely be improved by using Gaia DR3 and the above-mentioned supportive observational data, especially for the particularly interesting cluster containing $\beta$\,Lyrae itself.

\begin{acknowledgements}
      This work is based on data from the European Space Agency (ESA) mission {\it Gaia} (\url{https://www.cosmos.esa.int/gaia}), processed by the {\it Gaia} Data Processing and Analysis Consortium (DPAC, \url{https://www.cosmos.esa.int/web/gaia/dpac/consortium}). Funding for the DPAC has been provided by national institutions, in particular the institutions participating in the {\it Gaia} Multilateral Agreement.\\
      This research has made use of the SIMBAD database, operated by the CDS at Strasbourg, France. \\
      The author thanks Wolfgang Quester for pointing him to the old question on possible stellar partners of $\beta$~Lyrae.\\
      Furthermore, many thanks go to an anonymous referee. He/she made a few suggestions which definitely improved the paper.\\
       
\end{acknowledgements}


\bibliographystyle{aa} 
\bibliography{betLyr} 

\end{document}